\documentclass[twocolumn,prl,aps]{revtex4}
\usepackage{amsmath}
\usepackage{amssymb}
\usepackage{graphicx}
\usepackage{dcolumn}
\usepackage{bm}
\usepackage{epstopdf}

\begin{document}

\title{Topological insulators based on HgTe}

\author{Z. D. Kvon,$^{1,2}$ D. A. Kozlov,$^1$ E. B. Olshanetsky,$^1$ G. M. Gusev,$^3$ N. N. Mikhailov$^1$ and S. A. Dvoretsky$^1$}

\affiliation{$^1$Institute of Semiconductor Physics, Novosibirsk
630090, Russia}
\affiliation{$^2$ Novosibirsk State University, Novosibirsk 630090, Russia}
\affiliation{$^3$Instituto de F\'{\i}sica da Universidade de S\~ao
Paulo, 135960-170, S\~ao Paulo, SP, Brazil}

\begin{abstract}

The most interesting experimental results obtained in studies of 2D and 3D topological insulators (TIs) based on HgTe quantum wells and films are reviewed. In the case of 2D TIs, these include the observation of nonlocal ballistic and diffusion transport, the magnetic breakdown of 2D TIs, and an anomalous temperature dependence of edge-channel resistance. In 3D TIs, a record-setting high mobility (up to $5\times10^{5} cm^{2}V^{-1} s^{-1}$) of surface two-dimensional Dirac fermions (DFs) has been attained. This enabled determining all the main TI parameters (the bulk gap and the density of DFs on both of its surfaces) and provided information on the phase of the Shubnikov–de Haas oscillations of DFs, which indicates the rigid topological coupling between the fermion spin and momentum. Prospects for further research are discussed in the conclusion.
\end{abstract}

\maketitle
\section{Introduction}

The topological insulator (TI) is a concept that has appeared in condensed matter physics relatively recently, in 2007. And yet, several thousand studies pertaining to some extent to TIs have already appeared, and studies of TIs can be regarded, without much exaggeration, as one of the most actively developing areas of modern condensed matter physics. This is evidenced by the emergence of numerous reviews, among which the first two [1, 2] and several recently published ones [3–7] should be distinguished. However, the literature on TIs is clearly biased towards theory and model calculations, which creates a distorted impression of the actual situation in the physics of TIs.

A paradoxical situation in the exploration of 2D TIs took shape as a result of this imbalance: the most interesting transport property of these insulators—ballistic transport along the edge current states on a scale of several micrometers—has only been observed in experiments of the Molenkamp group at the University of Würzburg more than ten years ago [8–10], and until recently it had not been confirmed by any other group, including the authors of the original experiments [8–10]. Such confirmation has become possible only recently [11, 12]. Thus, theoretical studies of 2D TIs, which number several hundred, are based, essentially, on publications [8–10] alone.

The existence of diffusive regional transport, in contrast to ballistic transport, has been confirmed by other groups [13–15]. Ballistic edge transport in a quantum well based on the GaSb/InAs heterojunction was observed recently [16]; however, additional experiments are required to reach an unambiguous conclusion regarding the existence of a 2D TI in this system.

Thus, HgTe-based quantum wells (QWs) with an inverse spectrum remain essentially the only system in which the existence of a 2D TI has been reliably established; Sections 2–4 are devoted to the presentation of the physics of this TI. The bias mentioned above is even more prominent in the study of 3D TI. Most experiments explore 3D TIs based primarily on bismuth compounds (BiTe, BiSe, $Bi_2Te_2Se$, etc.), and all these studies (see [1, 2]) focus primarily on experiments on angle-resolved photoemission spectroscopy (ARPES), which has provided more than comprehensive information about the energy spectrum of surface electrons.

This information unambiguously evidences the existence of a whole set of materials whose surface is populated with charge carriers that have a linear Dirac spectrum and a rigid coupling of spin and momentum. However, because of the poor quality (the concentration of residual impurities is higher than $10^{17} cm^{-3}$) and low mobility (as low as about $10^3 cm^2/Vs$) of these materials, it is not possible to obtain the most interesting information related primarily to the transport response of Dirac surface electrons. In particular, experiments are still lacking in which the state of 3D TI would be realized where the Fermi level is located in the bulk gap and the transport response of the TI is only caused by surface Dirac states and not distorted by the bulk contribution. Attempts [17, 18] to solve this problem by drastically reducing the thickness of the samples (to 10 nm) resulted in a situation where the sample volume can no longer be considered three-dimensional, and the discussion of a 3D TI loses its meaning. For this reason, numerous ambitious predictions regarding exotic properties of TIs remain the domain of intellectual speculation, rather than interesting and profound physical exploration.

The situation with experimental studies of 3D TIs changed with the implementation of TIs based on strained HgTe films [19–21]. Studies of such TIs and the results obtained are described in Section 4.
2. Topological insulators. Background information

The most important property of all TIs is the presence of a delocalized band of surface states. We note that the emergence of such bands was discussed as early as the 1950s, in particular in review [22] devoted to Tamm and Shockley states. However, the first well-grounded calculations were done in the pioneering studies [23–25], which showed for the first time that the presence of spin–orbit coupling leads to the emergence of surface states, in particular, on the surface of mercury telluride and at the boundaries of a QW created on its basis. Several more papers appeared later in which this issue was discussed in relation to the valence band [26]. The results of the early studies were summarized in [27], where the exact Kane Hamiltonian was used to calculate the band spectrum of QWs based on HgTe and establish all the basic properties of dimensional quantization in such QWs, including the interaction between bulk and surface states and their mutual transformations. Of particular note are studies [28, 29], which were the first to indicate the possible emergence of surface bands of massless Dirac fermions (DFs) at the interface of semiconductors with inverse and direct spectra. However, none of these results has been reproduced in experiments, due to the lack of technology to obtain the required quantum wells.

A burst of research in TIs occurred when new theoretical ideas [30–33] were put forward that were almost immediately confirmed by experiment. In some ways (such as the emergence of boundary states), they iterated the ideas suggested in earlier studies, but, more importantly for the boom to develop, they showed that all these states can be unified on the basis of the universal idea of topological order, for which almost immediately an exact brand was coined—a topological insulator [34]. This also contributed to a large extent to the emergence of the topological boom.

We now discuss the concept of topological order in more detail. The idea is to introduce a $Z_2$ topological invariant that is expressed as an integral over the boundary of the bulk Brillouin zone [30, 31] and actually reflects a direct relation between the bulk and the surface. In the case of a normal insulator, $Z_2 = 0$, and for a TI, $Z_2 = 1$. In other words, $Z_2$ is equal to the number of topological zones on the surface.

Generally speaking, a similar topological approach had been developed in the analysis of the quantum Hall effect (QHE) long before the topological boom [35–37]. It is not without reason that the QHE-regime system is now cited as an example of a 2D TI. The $Z_2$ invariants can be constructed in mathematically various ways, but their physical meaning is directly related to the wave function symmetry, which changes radically as a result of the band spectrum inversion. Such an inversion is in fact due to the relativistic terms in the Hamiltonian of a crystal consisting of heavy atoms, such as Hg or Bi. The main contribution comes from two terms: the more significant term is due to the spin–orbit coupling, and the less important one, to the Darwin term.

There are three types of spectrum inversions: s–p, p–p, and d–f [38]. Distinguished in this series is mercury telluride, in which, as has long been known, the simplest type of s–p inversion is realized, in which the hole-like $\Gamma_8$ band lies 0.35 eV above the electron-like $\Gamma_6$ band. However, despite the spectrum inversion, bulk HgTe is not a topological insulator because a gapless state is realized in its bulk, which can only be broken by lowering the initial crystal symmetry by an external effect, of which uniaxial compression is an example [39].

A special situation is realized in QWs based on HgTe, where, as a result of dimensional quantization for the QW thicknesses above a critical value dc lying in the range 6.3–6.5 nm, an inverse gap emerges already in a 2D volume, and edge states emerge at the well boundaries, and thus a 2D TI is realized, with which we begin the presentation in Section 3.

\section{Two-dimensional topological insulator in an HgTe-based quantum well}
\subsection{Energy spectrum of HgTe quantum wells}

First, we describe the energy spectrum of a QW based on mercury telluride in more detail. Figure 1 shows a qualitative view of the bottom energies of the main dimensional-quantization subbands in such a well as a function of its thickness. As can be seen, the behavior of the spectrum fundamentally depends on the well thickness, and it can be conventionally divided into three regions. The first region is $d_w > d_c$, where a direct-band-gap 2D insulator is realized. Its band gap decreases as the thickness increases, to collapse at a critical well thickness dc, which is equal to 6.3–6.5 nm, depending on the QW orientation and deformation. As $d_w$ increases further, the second region appears, which contains a 2D insulator but with inverse bands. Finally, if $d_w >  15–16 nm$, a semimetal state [40, 41] emerges due to the overlap of hole-like bands H1 (conduction band) and H2 (valence band). Because the discussion in what follows is focused on the properties of 2D TIs, of interest for us is only the second region, in which a 2D TI is realized. The energy spectrum of this TI calculated in [42] for (100) and (013) surfaces is shown in Fig. 2a. As can be seen, the basic characteristics of this spectrum only weakly depend on the orientation of the surface. The critical thickness is in both cases $d_c = 6.2–6.3 nm$, and the 2D TI state with the largest band gap, which is characterized by the simplest s–p inversion, is realized at a QW thickness of 8.2–8.5 nm. The band gap width is in this case approximately 30 meV.

\begin{figure}[ht]
\includegraphics[width=8cm]{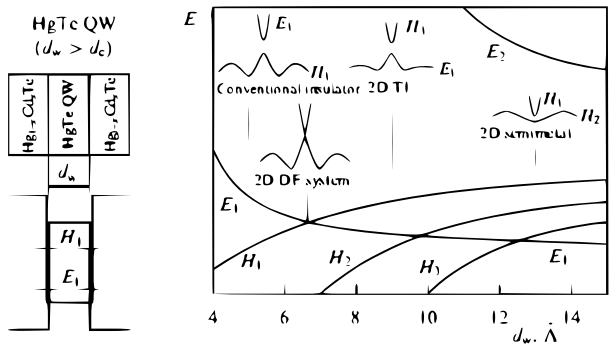}
\caption{(Color online) Qualitative view of the dependence of the subband bottom energy (E1 and E2 are the energies of the electron subband bottom, H1, H2, and H3 are the energies of the hole subband bottom) of dimensional quantization of the HgTe QW on its thickness dw.}
\end{figure}

\begin{figure}[ht]
\includegraphics[width=8cm]{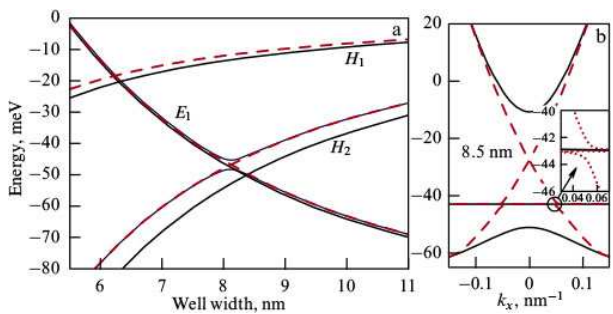}
\caption{(Color online) (a) Subband bottom energy as a function of the QW thickness in the range 5.5–11 nm (solid curves are for surface orientation (100), dashed curves for orientation (013)). (b) Dispersion law of bulk and edge states for HgTe QWs with orientation (013). (The figure is taken from [42].}
\end{figure}

The dispersion law for edge and bulk states for a QW with the thickness 8.5 nm and orientation (013) is shown in Fig. 2b, which well illustrates all the features of the spectrum of a 2D TI based on the HgTe QW: the linear Dirac spectrum of edge current states and a parabolic band-gap spectrum of bulk states. We note that the edge states exist not only in the band gap but also at energies that correspond to the allowed bulk bands. Figure 2b also clearly shows the anticrossing of the edge states with the bulk states in the lower part of the band gap due to lower surface symmetry (013).

\subsection{Experimental samples. Field-effect transistor based on a quantum HgTe well}

The vast majority of experimental samples used in the studies reviewed here are made on the basis of QWs with a given thickness of 8 or 8.3 nm and orientation (013). This orientation is chosen because, on the one hand, the presence of steps on suchlike surfaces ensures a more equilibrium growth of the HgTe and HgCdTe layers, which reduces the concentration of various point and dislocation defects, and, on the other hand, as shown in Fig. 2, the energy spectrum of a 2D TI does not substantially depend on orientation.

It is also of importance that in specifying the QW thickness, its exact value for a given sample may not correspond to the specified growth thickness, and deviations from it by several tenths of a nanometer are possible due to the heterogeneity of the atomic beam density in the process of molecular beam epitaxy.

The QW itself does not enable a full-fledged study of the state of a specifically 2D TI, because two conditions must be satisfied for its transport response to be observable: it is necessary to ensure that the Fermi level is located in the bulk band gap, and the clearest, most reliable, and simplest way to detect edge states is needed. The first condition may be satisfied using the field-effect transistor structure schematically shown in Fig. 3a. The second condition is specifically discussed in detail in Section 3.3.

\begin{figure}[ht]
\includegraphics[width=8cm]{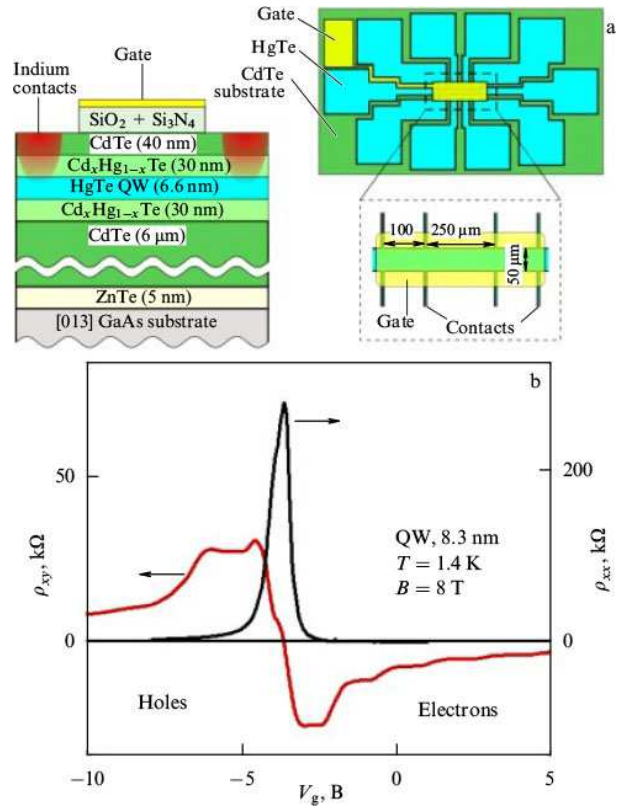}
\caption{(Color online) (a) Transistor Hall structure based on HgTe QW and HgTe films and (b) dependences, typical of this structure, of the dissipative and Hall resistance components on the gate voltage for a QW 8 nm in thickness.}
\end{figure}

Two more operations are required to produce a field-effect transistor based on the HgTe QW: low-temperature growth of the dielectric layer and the subsequent deposition of a metal gate on it. Either a pyrolytic $SiO_2$ layer or a double $SiO_2 + Si_3N_4$ layer grown at temperatures of $80–100^{\circ}C$ was used as a dielectric, and the Ti/Au layer served as a gate. We note that there are other methods to grow dielectric layers, but we do not discuss them here.

We now briefly describe the conditions under which the transport response of the investigated 2D TI was measured. The measurements were carried out in the temperature range 0.2–10 K in magnetic fields up to 15 T using the standard phase-sensitive detection scheme at frequencies of 6–12 Hz and currents 0.01–10 nA, depending on the type of experiment, to exclude the effects of heating of the electronic subsystem.

Figure 3 shows the dependences of the dissipative and Hall components of the resistance tensor on the gate voltage, typical of a macroscopic sample made on the basis of an 8 nm HgTe QW. It is clearly seen that the resistance is small (of the order of $100 \Omega/\Box$) at displacements that correspond to the location of the Fermi level (EF) in the conduction band, passes through a maximum (equal to about $300 k\Omega$ in this case) that corresponds to $E_F$ occurring in the middle of the bulk gap, and then begins to decrease, reaching values of several $k\Omega/\Box$, when the Fermi level enters the valence band. The point of the maximum of $\rho_{xx}$ is commonly referred to as the charge neutrality point (CNP). The dependence $\rho_{xy}(V_g)$ in turn exhibits a well-pronounced plateau at the Landau level filling factors i = 1 and i = 2 on the electron side, passes through zero at the CNP, and has the opposite sign in the valence band, but the plateaus are no longer observed due to significantly lower (by an order of magnitude) hole mobility. The absence of the Hall signal at the CNP indicates that there are no mobile charges in the QW. Thus, this point fully justifies its name.

We note that strictly speaking, a zero of the Hall signal and especially the maximum of resistance are not direct evidence of the absence of charge in the well. Therefore, caution is needed in every determination and analysis of the CNP. Another feature of $\rho_{xy}(V_g)$ curves is of importance: their half-width is significant (about 1 V). This feature indicates that the density of states inside the bulk band gap is quite high, a feature that has been overlooked in the vast majority of studies on 2D TIs. This is discussed in more detail in Section 3.4.

\subsection{Experiment. Detection of edge current states}

The dependence shown in Fig. 3b essentially says nothing about edge transport, because the measurement does not allow eliminating the influence of the bulk. Of crucial importance for determining edge transport is measuring it in a nonlocal geometry.

We now make some preliminary remarks regarding the essence of the resistance of the 2D TI edge channel. We compare this resistance with that measured in the QHE regime, which is also a kind of 2D TI.

We consider the simplest example: a two-terminal conductor of length L and width W with ohmic contacts L (Left) and R (Right) in the cases where transport through it is maintained by the 2D TI edge states in the ballistic mode (Fig. 4a) or in the QHE mode in filling the degenerate Landau ground level (Fig. 4b). We start with the first option. The current transport is maintained in this case by two single-mode quantum wires with removed spin degeneracy, located on the lower and upper edges of the conductor. The state that bears the electrochemical potential of the left contact $\mu_L$ is located on the same edge of the sample where the oppositely directed state having the potential $\mu_{R}$  is localized.

\begin{figure}[ht]
\includegraphics[width=8cm]{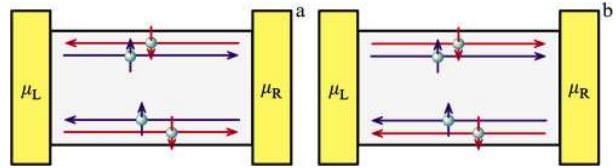}
\caption{(Color online)(a) Two-dimensional topological insulator. (b) Two-dimensional conductor in the QHE regime when the ground Landau level alone is filled. }
\end{figure}

Exactly the same reasoning applies to the opposite edge of the sample, and we therefore actually have two single-mode ballistic wires connected in parallel, each of which has a conductance equal to $e^2/h$. Then the measured conductance $G_{12, 12} = eI_{12}/(\mu_{L}- \mu_{L})$ is equal to $2e^2/h$. The situation in the QHE regime is completely different: the current is transported by spatially separated states, i.e., those states localized at opposite edges of the conductor. One of them bears the electrochemical potential of the left contact $\mu_{L}$, and the other, of the right contact, $\mu_{R}$, and the measured conductance has the same magnitude $2e^2/h$.

We now consider a 2D conductor with a resistivity $\rho_{xx}$ of length L and width W with contacts 1–4, as shown in Fig. 5. If the current passes through contacts 1 and 2, and the voltage is measured at contacts 3 and 4, the resistance $R_{1234} = V_{34}/I_{12}$ is by the order of magnitude equal to [43].

$R_{1234}\approx\rho_{xx}exp\left(-\frac{\pi}{W}\right)$

As the ratio of the conductor length to its width increases, the resistance decreases exponentially for a trivial reason: only an exponentially small part of the total current reaches contacts 3 and 4. This is the configuration that corresponds to the measurement of nonlocal resistance.

\begin{figure}[ht]
\includegraphics[width=8cm]{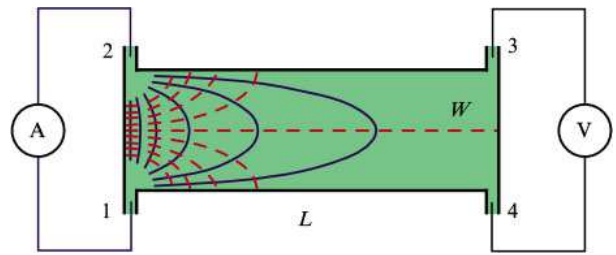}
\caption{(Color online)(a) Two-dimensional conductor and distribution of currents in it. }
\end{figure}

We now consider the situation where a band gap emerges in the bulk of the conductor. No current flows through it in the case of a conventional insulator. However, if it is a TI, the entire current flows through the edge states, because they are delocalized. In the case of ballistic transport, we then obtain $R_{1234}=\frac{h}{4e^{2}}$

and, in the case of diffusive transport, a weak linear decrease of the resistance, $R_{1234}=R_{L}\frac{W^{2}}{L+W}$

where $R_{L}$ is the edge-wire resistivity. It is hence apparent that a comparative analysis of the local and nonlocal responses enables unambiguous determination of the presence of edge transport and hence the edge states that make its occurrence possible. An example displayed in Figure 6 shows a typical measurement of local ($R_{loc}$) and nonlocal ($R_{nonloc}$) resistances for a sample (whose topology is shown in the inset in Fig. 6) made on the basis of an HgTe QW 8 nm in thickness. At first glance, the behavior of these resistances is qualitatively similar and coincides with that of the $\rho_{xx}(V_g)$ dependence shown in Fig. 3. But a more careful comparative analysis of $R_{loc}$ and $R_{nonloc}$  reveals a significant difference between their behaviors: while the $R_{loc}$ values are sufficiently large at all gate voltages, including those that correspond to the location of the Fermi level in the allowed band, $R_{nonloc}$, as it should be, is close to zero at the specified voltages. However, $R_{nonloc}$ becomes comparable to Rloc in the vicinity of the CNP, i.e., when the Fermi level is located in the center of the bulk gap. Just this property of the 2D TI transport is a direct indication of the existence of charge transport along the edge.

\begin{figure}[ht]
\includegraphics[width=8cm]{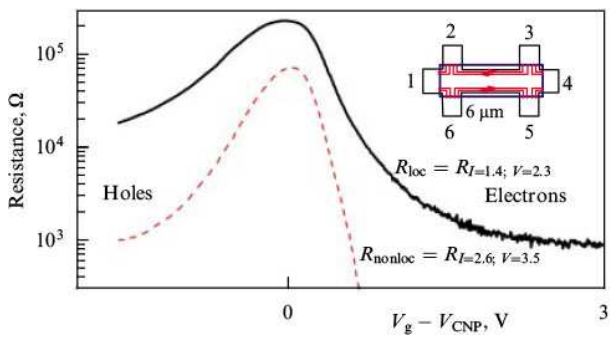}
\caption{(Color online)(a)  Local and nonlocal resistance in the diffusive regime. }
\end{figure}

The first experiments on edge transport were carried out in [8–10], where ballistic edge transport in an HgTe QW with a thickness of 7–8 nm was demonstrated using submicron-size samples. It was next shown in [11] that transport along edge states in these QWs also exists on macroscopic scales, of the order of 1 mm, but this time in the diffusive regime. The experiments in [44, 45] should also be noted, in which edge states were visualized, thus confirming their presence.
\subsection{Description of edge transport in a 2D topological insulator on the basis of a network model}

The easiest way to analyze edge transport is to use the Kirchhoff relations under the assumption that there is no bulk conductivity. An equivalent circuit in the case of a standard Hall bridge with two current and four potential contacts is shown in Fig. 7. It is easy to see that the resistance between contacts i and j can be expressed in terms of the resistance between all other contacts or in terms of the distance between them (the latter reflects the proportionality of the resistance of the edge wire to its length): $R_{m,n}^{i,j}=\frac{h}{e^{2}}\frac{L_{n,m}L_{i,j}}{lL}$

where the current flows through the contacts i and j, and $L_{i, j} (L_{m, n}$) denote the length of the edge states located under the metal gate that do not include these contacts, and L and l are the edge state circumference and the mean free path, respectively. This simple formula enables predicting the relation between local and nonlocal resistances in any configuration, which no longer depends on the mean free path and therefore on the presence or absence of backscattering. Formula (1) in the ballistic transport limit is the Landauer–Büttiker relation. The resistances of ballistic 6- and 4-pin bridges calculated in this way are in good agreement with experimental data. A comparison of the results calculated using Eqn (1) with measurements of samples exhibiting diffusion transport revealed a significant disagreement, which increases as the edge state length increases [13]. It is natural to assume that if the regime deviates from the ballistic one, it is necessary to take not only backscattering between the edge states but also their scattering into the bulk into account. The Kirchhoff rules are not suitable for a quantitative analysis of such a deviation. The problem can be resolved by introducing two phenomenological parameters $\gamma$ and $g$ for the rate of scattering between the edge states and the edge state and the bulk [46].

\begin{figure}[ht]
\includegraphics[width=8cm]{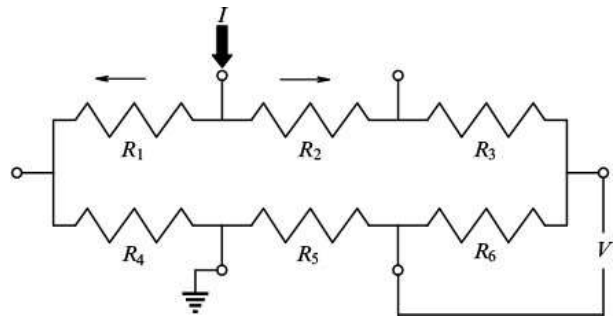}
\caption{(Color online)(a)Equivalent circuit of a 2D TI in the edge transport regime in the case of a standard Hall bridge with two current and four potentiometric contacts. The edge channel is replaced with the equivalent resistance $R_k = (h/e^2)(L_k/l)$, where $L_k$ and l are the channel length and mean free path. Current inflows between the resistances R1 and R2 and outflows between R4 and R5; the voltage drop is measured on the resistance R6.  }
\end{figure}

We recall that different edge states running towards each other are associated with different spins. The distribution of the bulk potential is determined in this case using the Laplace equation and the corresponding boundary conditions. The distribution of the edge state potential is found using the balance equation [47]. Both the edge and bulk potentials that belong to different spin states are mixed in the contact region, which in our case is a 2D electron gas outside the gate area (which covers only the central part of the Hall bridge).

It is of importance that the results calculated in this model are independent of the mechanisms of microscopic scattering between the edge states or the edge channels and the bulk. Possible scattering mechanisms are described below.

A specific feature of the model under consideration is that transport in various systems is described in the model in a universal way. This model was applied first to a 2D system demonstrating the QHE, when the chiral edge state that belongs to the last Landau level is mixed with the bulk level, resulting in a significant nonlocal response [47]. The model also successfully described the quantum transport of the zeroth Landau level of DFs in graphene, which form edge modes running towards each other [48]. The bulk transport can be described in both of these cases as the transport of a 2D electron in a quantizing magnetic field. Edge transport occurs in a TI when the bulk is an insulator in the absence of a magnetic field, and describing bulk transport requires a different approach. It has been suggested that bulk transport is determined by Gaussian tails of the density of states due to the presence of a fluctuation potential arising from fluctuations in the QW thickness and an impurity potential. Based on this assumption, a description was provided for both local and nonlocal transport in the presence of scattering both between the edge states running towards each other and between the edge states and the bulk using $\gamma$ and $g$ as fitting parameters [46].

The Gaussian broadening of the density of states was found in this case using the mobility of electrons and holes at the bottom of the corresponding bands. The proposed model, which takes the leakage of current into the bulk into account, describes the dependence of the resistance on the density of charge carriers fairly well (Fig. 8). Indeed, if the Fermi level is located in the center of the forbidden band, current leakages through the bulk are minimal, and the resistance is determined by the edge transport, i.e., scattering between edge states. As the Fermi level approaches the valence or conduction band, the contribution of the bulk due to the scattering of edge states into the bulk, as well as the contribution due to an increase in the bulk conductivity itself, increases, and the total resistance decreases. The width of the resistance peak is then determined by the velocity of the Fermi level motion through the tails of the density of states in the topological insulator band gap, and, to describe the peak width observed in the experiment, it is necessary to assume a high density of states inside the band gap, only several times lower than the bulk one.

\begin{figure}[ht]
\includegraphics[width=8cm]{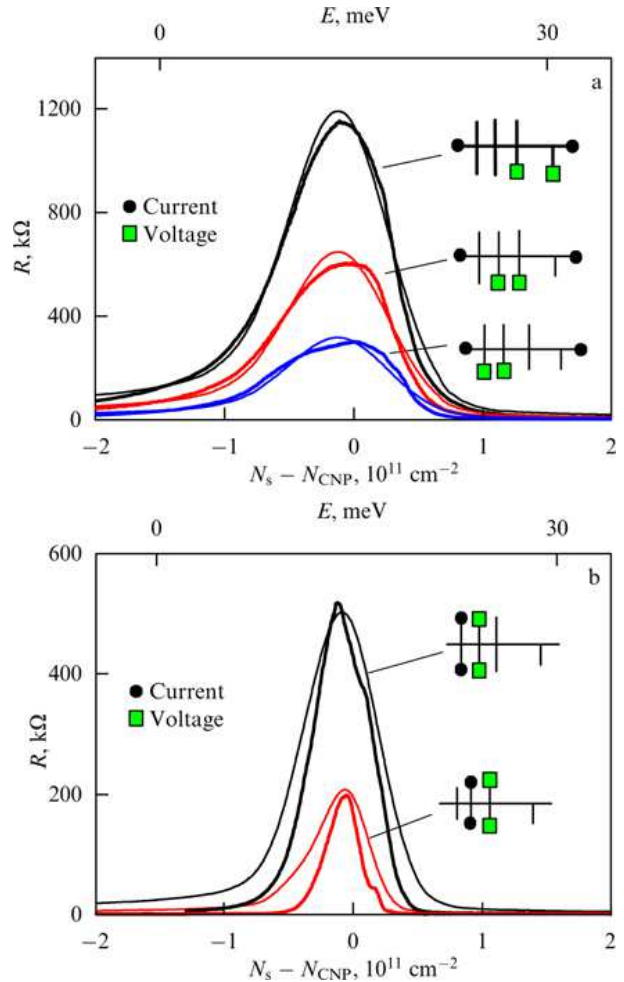}
\caption{(Color online)Comparison of the dependences of (a) local and (b) nonlocal resistances on the density of charge carriers with the results of a model that takes edge and bulk scattering into account for various configurations of applied current and measured voltage. The calculated dependences are shown by thinner lines. The calculation parameters are specified in [46].
  }
\end{figure}

\subsection{ Temperature dependence of the resistance of a 2D topological insulator}

In this section, we analyze measurements of the temperature dependence of the resistance of a 2D TI in the diffusive transport mode. Such measurements are of importance for two reasons: first, it is necessary to determine the activation band gap and compare its value with the calculated one, while the second, more fundamental, reason is due to the edge channel of a 2D TI being an almost perfect 1D conductor, a fact that at first glance allows testing all predictions of numerous theories of 1D conductivity.

Typical measured temperature dependences at high (T > 4 K) temperatures for the samples described in Sections 3.1–3.4 are displayed in Fig. 9. The inset in Fig. 9a shows the topology of the samples. It is clearly seen that at temperatures higher than T = 30 K, an activation increase in resistance is observed, which is followed by complete saturation of the R(T) dependence, and, if the temperature decreases to 4 K, no significant temperature dependence is observed.

\begin{figure}[ht]
\includegraphics[width=8cm]{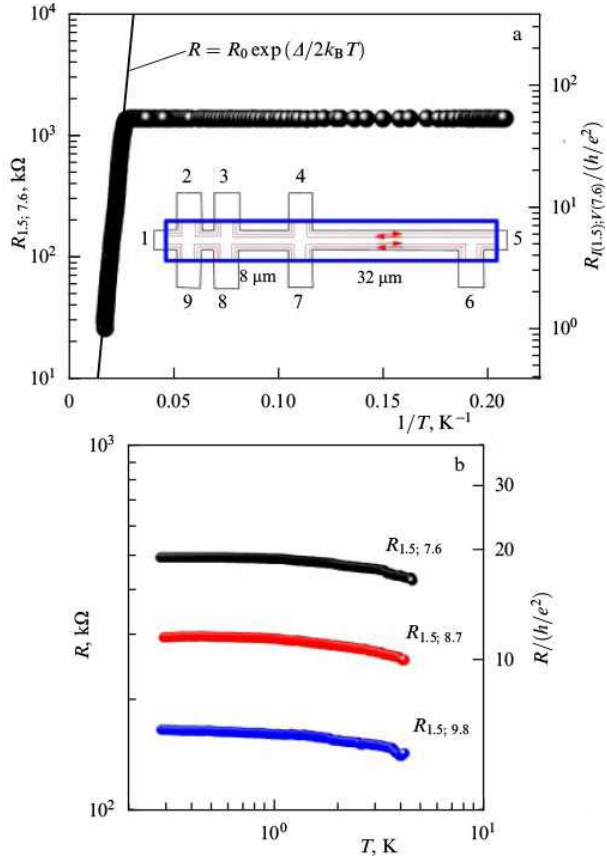}
\caption{(Color online)Temperature dependence of the resistance of a sample based on an HgTe QW 8 nm in thickness at (a) high ($100 > T > 4 K$) and (b) low ($4 > T > 0.2 K$) temperatures.}
\end{figure}

The activation increase in resistance is associated with the freezing of bulk conductivity, the activation energy for the dependence shown in Fig. 9 being approximately 200 K. We note that the activation energy can vary significantly, depending on the sample, in the range 200–400 K. Calculations of the energy spectrum for the wells under study shown in Fig. 2 yield a gap of about 30 meV, which fits into the specified activation energy range. We note, however, that suchlike measurements actually determine the mobility band gap, which heavily depends on disorder, whose significant role is suggested by the large spread of the activation energy.

At temperatures above 20–30 K, the temperature dependence of the resistance of the 2D TI based on HgTe QWs with a thickness of 8–9 nm in the regime of diffusive transport reflects the bulk properties, more precisely, the size of the mobility band gap in the bulk. As was noted above, as the temperature decreases further to 4–5 K, the resistance ceases to change and, when transport is actually maintained by edge states, a metallic behavior of the resistance is observed. To check whether this (metallic) behavior persists at lower temperatures, measurements were carried out using a dilution refrigerator at temperatures down to 40 mK [49]. Figure 9b, where the results of these measurements are displayed, shows that the character of the temperature dependence barely changes: in the temperature range 4–1 K, only a very weak ($10\%$) increase in resistance is observed, and further, at temperatures down to 40 mK, there is no temperature dependence whatsoever.

To date, various approaches have been proposed to explain this behavior of resistance, but so far all of them have failed to provide a comprehensive explanation. The model of metal droplets proposed in [50, 51] seems to be preferable. According to this model, an electron moving along an edge state enters such a droplet, and backscattering can occur inside it as a result of inelastic scattering, which leads to suppression of ballistic transport and to the conductance values lower than the conductivity quantum. However, the temperature dependence predicted by this model disagrees with the experimental data: as the temperature decreases, the edge conductance should increase due to the suppression of inelastic processes, and at temperatures tending to zero, a conductance close to $e^2/h$ should be observed, while in the experiment it is practically independent of temperature in the range from 20 to 0.2 K. Perhaps this shows that along with impurity disorder, the structural disorder caused by fluctuations in the QW thickness are to be taken into account.

\subsection{ Magnetotransport properties of edge current states}

We first describe the response of edge transport in a 2D TI to a normal magnetic field. Figure 10a shows a typical dependence of local and nonlocal resistances on the magnetic field at the CNP. It can be seen that the qualitative behavior of the two dependences is the same, i.e., in both local and nonlocal geometry, positive magnetoresistance (MR) is observed in magnetic fields up to 2 T, which is followed by its rapid decrease and subsequent equally rapid growth. The displayed results thus clearly show the edge nature of the MR. Figure 10b shows the behavior of linear magnetoconductivity (MC) in more detail at temperatures of 4.2 and 1.6 K (data of local measurements are quoted in this case). It can be seen that MC weakly depends on temperature and is about $10–15\%$ in magnetic fields of about 1 T. This linear MC behavior was predicted in [52], where it was explained by the suppression of topological protection from back scattering by the magnetic field.

\begin{figure}[ht]
\includegraphics[width=8cm]{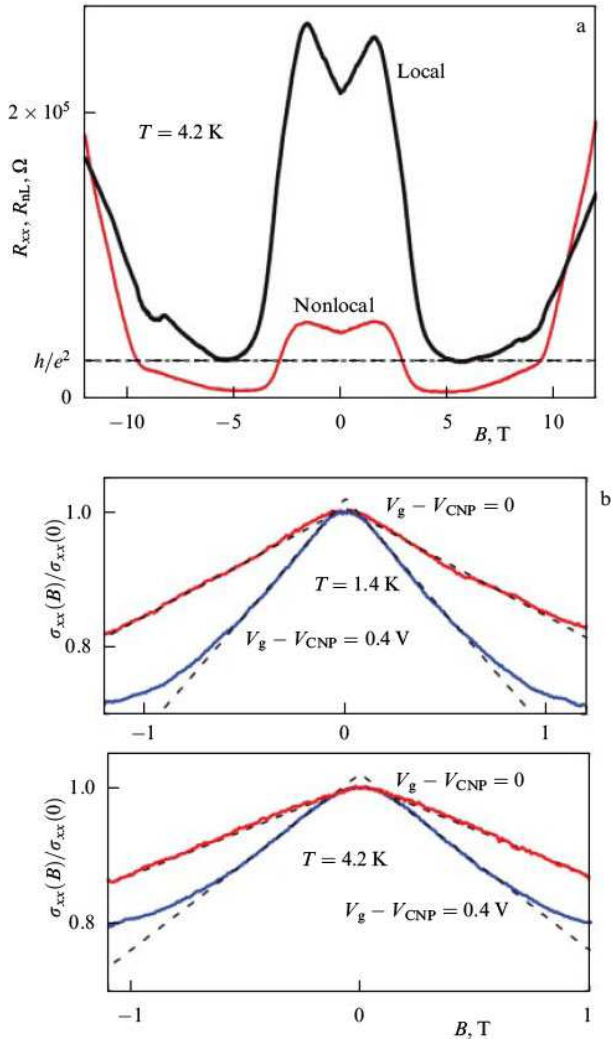}
\caption{(Color online)(a) Magnetoresistance of a 2D TI based on an 8 nm HgTe QW in local and nonlocal geometry. (b) Normalized magnetoconductivity as a function of the magnetic field in the range $|B|<1.2 T$. Dashed curves show linear approximations of the dependences.}
\end{figure}

We next consider the effect of the magnetic field in the QW plane. Figure 11a, b shows local and nonlocal resistance of the sample (whose topology is shown in the inset in Fig. 11a) as a function of the magnetic field. It is clearly seen that the same behavior is observed at all temperatures: if the magnetic field is less than 8 T, $R_{xx}$ and $R_{nL}$ monotonically decrease to values that are 1.5 to 2 times smaller than those in the zero magnetic field. The decrease then becomes faster and, at $B > 10 T$, $R_{xx}$ virtually saturates at a value that is already an order of magnitude smaller, while $R_{nL}$ is close to zero.

\begin{figure}[ht]
\includegraphics[width=8cm]{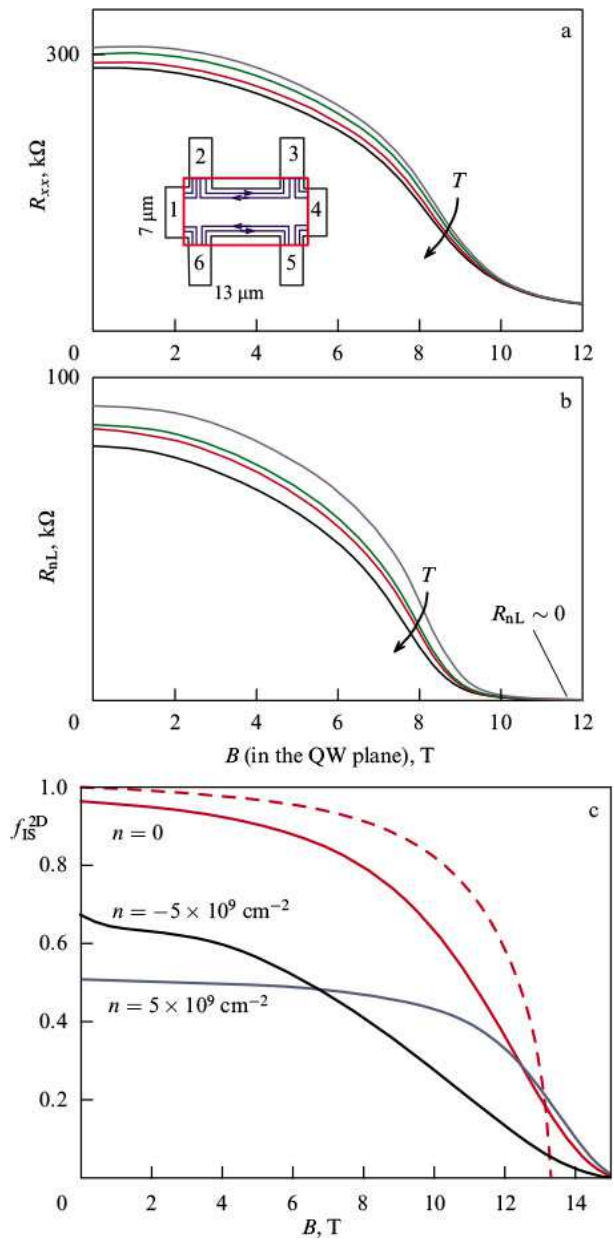}
\caption{(Color online) Local and nonlocal resistance of a 2D TI depending on the magnetic field applied in the QW plane: (a) local magnetoresistance, (b) nonlocal magnetoresistance, (c) theoretical calculation of the fraction of the dielectric state in the 2D plane with and without taking impurity disorder into account. }
\end{figure}

This result has been explained in the theory developed by Raichev [53]. According to this theory, if a magnetic field is applied in the QW plane, the bulk gap gradually decreases and, in magnetic fields stronger than 12 T, the bulk transforms into the state of a 2D Dirac semimetal with a corresponding emergence of bulk conductivity.

Figure 11c shows how the QW bulk transforms into a semimetal state at various degrees of disorder. Good agreement between theory and experiment is clearly seen both in the character of the dependence on the magnetic field and in its magnitude (10–11 T in the experiment and 12–14 T in calculations), which corresponds to the complete transition of the QW into a gapless state. A slight disagreement in the values of critical fields is not a surprise, because all parameters of the system (bulk parameters of HgTe and CdTe, impurity concentration, the QW thickness fluctuations) that are used in the calculations are only known with a certain accuracy.

\subsection{Terahertz photoresistance of a 2D topological insulator}

Here, we present the results of an experimental study of the terahertz photoresistive response of a 2D TI [12] in which the regimes of ballistic and quasiballistic transport were successfully implemented. The experimental samples were microstructures of a special Hall geometry, equipped with a semitransparent Ti/Au gate (Fig. 12a) whose characteristic dimensions are comparable to the mean free path along the edge state. In particular, the microbridge width was 3.2 $\mu m$.

\begin{figure}[ht]
\includegraphics[width=8cm]{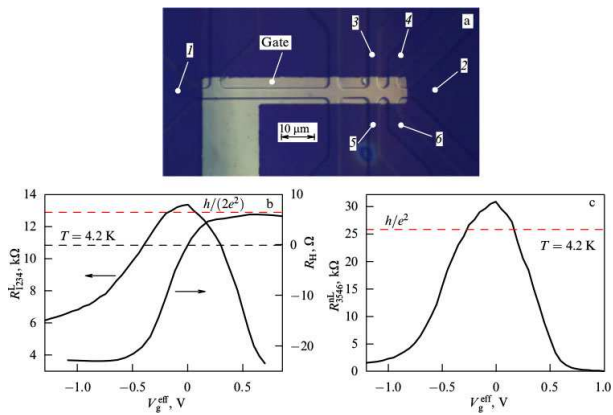}
\caption{(Color online) (a) Photo of the microstructure of the special Hall geometry. (b) Local resistance at B = 0 and the Hall resistance at B = 1 T as a function of the effective gate voltage . (c) Dependence of the nonlocal resistance at B = 0. }
\end{figure}

The terahertz resistive response (photoresistance) of the described structures was measured at a wavelength of 118 $\mu m$ in transverse and longitudinal magnetic fields up to 4 T at temperatures $T = 2–4.2 K$. A molecular submillimeter methanol-based laser with optical pumping by a $CO_2$ laser was used as a radiation source. The terahertz radiation power $P_{\lambda}$ was in the range 20–30 mW. The photoresistance (PR) was measured using a standard modulation technique at a modulation frequency of 600–700 Hz while passing the direct current I = 100 nA through the sample.

We begin the description of the experiment with analyzing the transport response of the samples studied. Figure 12b shows the dependences on the effective gate voltage ($ V_g$ is the gate voltage and is the gate voltage that corresponds to the maximum local resistance) of the Hall resistance and the local resistance measured on the shortest part of the sample, where the distance between the potentiometric contacts (contacts 3 and 4 in Fig. 12a) was 2.8 $\mu m$. As can be seen, the resistance is small (of the order of $1 k\Omega/\Box$) at displacements that correspond to the location of the Fermi level (EF) in the conduction band, passes through a maximum (equal to 13.4 $k\Omega$ in this case) at the CNP (EF simultaneously passes through the Dirac point), and then begins to decrease to reach several $k\Omega/\Box$ when the Fermi level enters the valence band. The dependence passes through zero and changes sign. Figure 12c shows the nonlocal resistance of the sample , when contacts 3–5 and 4–6 were used as respective current and potentiometric contacts. As expected, the signal of the nonlocal resistance is much smaller than that of the local one when the Fermi level is located in allowed bands. At the CNP, the nonlocal signal is almost three times higher than the local one.

We now analyze the quoted data. The local resistance at the maximum is close to $h/(2e^2)$ (shown by the dashed line in Fig. 12b). This implies that virtually ballistic transport is realized in the smallest section of the studied Hall structure (about $10 \mu m$ along the sample edge). We note that this was the first observation of such transport in QWs 8–9 nm thick after the publication of [8, 9]. The nonlocal resistance is determined by splitting the current passing through contacts 3–5 between the parts of the sample with ballistic transport and with diffusive transport. The value of therefore lies between $2h/e^2$ and $h/e^2$. The value of $h/e^2$ is shown in Fig. 1c by a straight dashed line.

Figure 13a displays typical measured local PR of the sample as a function of the gate voltage under the effect of about 20 mW terahertz radiation at a wavelength of 118 $\mu m$. For the convenience of comparative analysis, the same figure shows the dependence . The dependence of the nonlocal PR at the same power values is shown in Fig. 13b. The dependence is also displayed in the same figure.

\begin{figure}[ht]
\includegraphics[width=8cm]{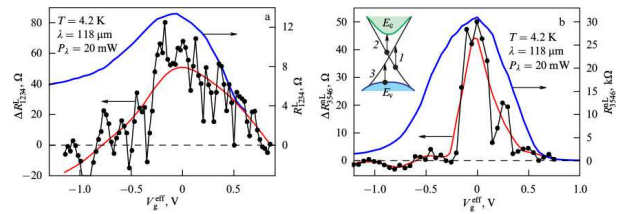}
\caption{(Color online)(a) Local photoresistance and the resistance as functions of the effective gate voltage. (b) Nonlocal photoresistance and the resistance as functions of the effective gate voltage. Red curves are shown to guide the eye. }
\end{figure}

We now discuss the quoted data. It is clearly seen that both local and nonlocal PRs are virtually zero when the Fermi level is located in allowed bands, to become nonzero only when the Fermi level enters the forbidden band, and the PR sign is negative, i.e., resistance of the sample decreases under the effect of radiation. Upon reaching the CNP, both dependences and pass through a maximum at which their value is $0.1–0.5\%$ of the total resistance. A more detailed comparative analysis of the curves presented in Fig. 13 shows that while the half-width of the local PR peak virtually coincides with that of the local resistance peak, the width of the PR dependence on in the nonlocal case is more than two times smaller than the same dependence for the resistance.

We next discuss the results obtained. The bulk gap for the 8 nm QWs considered in this section is 30 meV, i.e., several times higher than the photon energy for the employed wavelength of $118 \mu m (\hbar\omega = 10.8 meV)$. Three types of transitions are then possible in our case: (1) between Dirac branches of 1D edge states; (2) between the electron Dirac branch and the conduction band; and (3) between the valence band and the hole Dirac branch. Transitions of the last two types would apparently lead to the emergence of PR maxima near the allowed bands, i.e., to the right (for transitions of the second type) or to the left (for transitions of the third type) of the CNP on the PR dependences on . This behavior is not observed in experiment. Thus, only transitions of the first type remain. A preliminary analysis of the absorption at these transitions carried out in Ref. [54] showed that dipole transitions between edge Dirac branches are forbidden, and only significantly weaker magnetic dipole transitions occur.

However, it was shown recently in [55] that a similar conclusion is not valid for the HgTe QW because the argument does not include spatial inversion violation at the boundaries of these QWs with barrier HgCdTe layers. It was found in [55] that due to the violation of inversion symmetry at these boundaries, direct dipole transitions between the edge branches are allowed, and formulas for the absorption coefficient have been derived. Thus, the experimental conclusion on the possibility of direct dipole transitions between edge branches has been confirmed theoretically in [55].

\subsection{ Two-dimensional topological insulator with a complex bulk spectrum}

The existence of a 2D TI was discovered recently [11] in 14 nm QWs with (112) orientation. The TI state with ballistic transport at distances of the order of 10 $\mu m$ was obtained in this TI for the first time after [8, 9]. Therefore, we discuss the results of [11] in more detail.

A qualitative view of the spectrum for such a QW is shown in Fig. 14a. We note that the bulk spectrum is no longer as simple as for QWs with a thickness of 8–9 nm: two more branches of hole states emerge between the s and p states: hh2 and hh3. As a result, the band gap and edge states of importance for the experiment turn out to be located between two hole branches, hh1 and hh2, while the band gap is much smaller, about 3.3 meV.

\begin{figure}[ht]
\includegraphics[width=8cm]{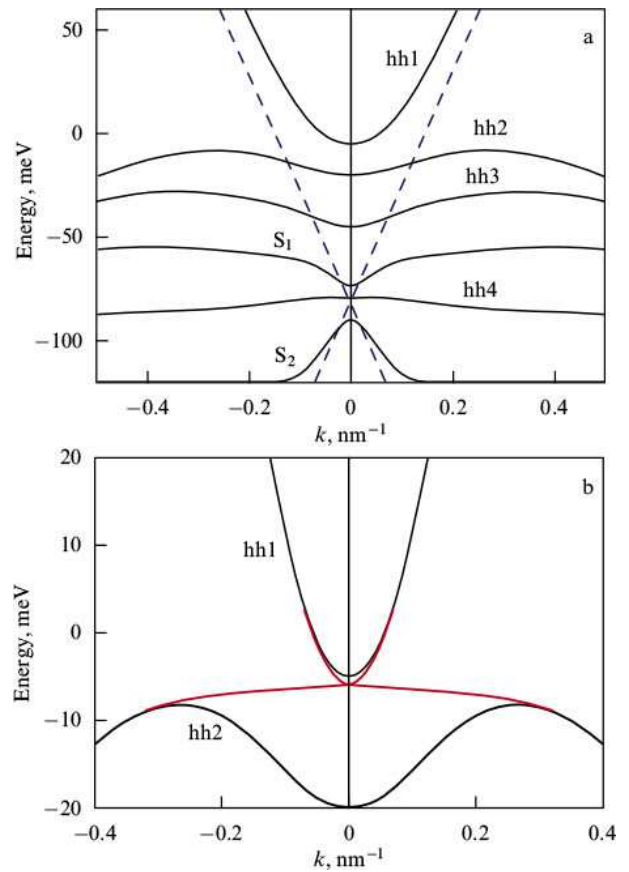}
\caption{(Color online)Qualitative view of the dispersion law of the bulk and edge states of 2D TI in a 14 nm HgTe QW: (a) spectrum without taking the mixing of bulk and edge states into account (S1 and S2 are the dispersion laws of electron subbands); (b) resulting spectrum for the two upper hole branches and edge states.}
\end{figure}

Activation measurements using macroscopic samples (Fig. 15a, b) yield a noticeably smaller gap (1.2 meV), which is unsurprising because the experiment based on measuring the activation temperature dependence actually determines the mobility band gap, which is always less than the real band gap due to disorder caused by the fluctuation potential of impurities and the QW composition and thickness.

\begin{figure}[ht]
\includegraphics[width=8cm]{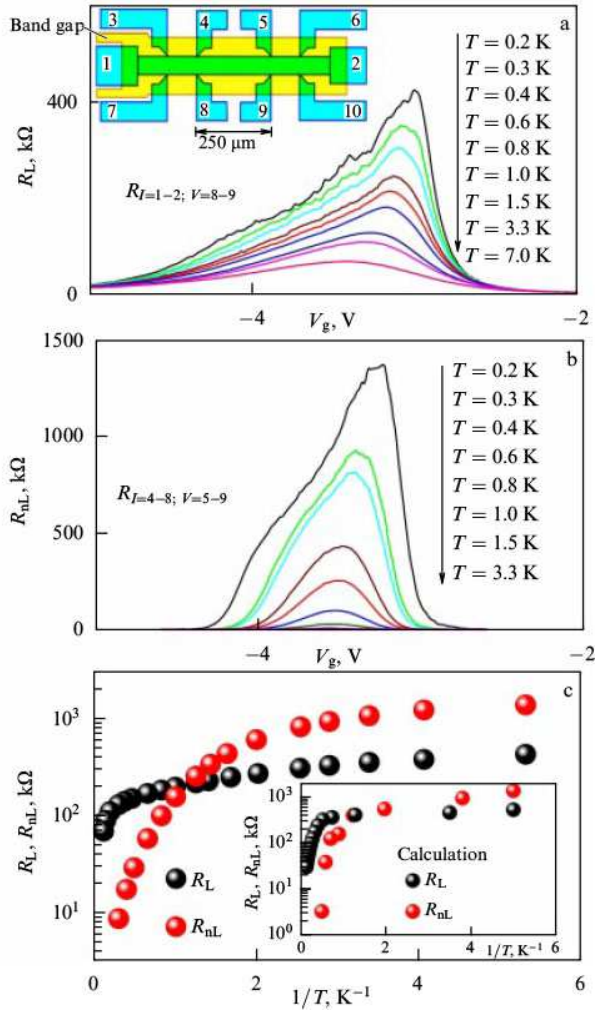}
\caption{(Color online)Temperature dependence of the edge transport of a macroscopic sample based on a 14 nm HgTe QW. (a) Local and (b) nonlocal resistance as a function of the gate voltage at various temperatures. (c) Temperature dependences of the local and nonlocal resistance at the CNP.}
\end{figure}

Figure 16 shows the results of experiments with micrometer-size samples. These results clearly demonstrate, on the one hand, the existence in such samples of both local and nonlocal transport close to the ballistic one, and, on the other hand, its apparent imperfection caused by mesoscopic fluctuations. The results presented are consistent in this regard with those obtained previously for QWs with a thickness of 7–8 nm [8, 9], where similar fluctuations were also observed. We note that ballistic transport exists in the described experiments in samples whose characteristic size is about $10 \mu m$.

\begin{figure}[ht]
\includegraphics[width=8cm]{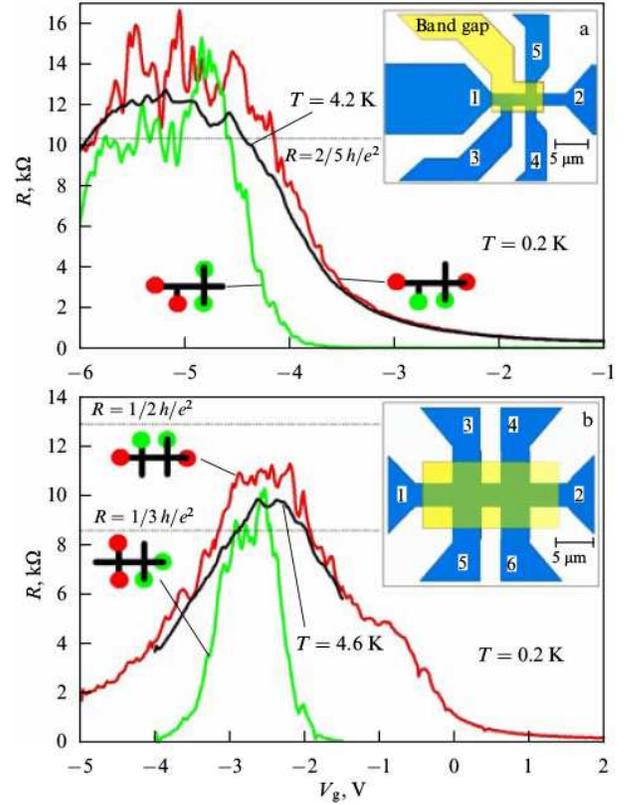}
\caption{(Color online)The measured local and nonlocal resistances of micrometer-size samples based on a 14 nm HgTe QW.}
\end{figure}

Thus, quasiballistic transport is possible in 2D TIs at distances that are significantly larger than the mean free path, which in the samples studied did not exceed $1 \mu m$ at energies close to the conduction band bottom.

\section{Three-dimensional topological insulator based on a strained HgTe film}
\subsection{Samples and experiment}

As noted in the Introduction, bulk mercury telluride, despite the inverse nature of its spectrum, cannot be classified as a TI because it is a gapless semiconductor.

However, if a uniaxial compression deformation is applied to HgTe, leading to the emergence of a band gap in the bulk, then a transformation into a 3D TI state can occur. A similar but not identical situation occurs, as shown in [19], for HgTe films grown on CdTe substrates due to the difference in the lattice constants of HgTe ($a_{HgTe} = 0.646 nm$) and CdTe ($a_{CdTe} = 0.648 nm$). The critical thickness of pseudomorphic growth that corresponds to this difference in the lattice constants is more than 100 nm, and thus films whose thickness is less than this critical value reproduce the lattice constant of the CdTe substrate. Tensile deformation occurs as a result in such films, leading to the emergence of a gap.

However, the Dirac point in the TI created in this way is located not inside the band gap but deep in the valence band. Due to hybridization with the valence band, the spectrum of surface states only contains the electron branch, which, when approaching the valence band bottom, deviates from the linear law to become quasiparabolic.

It can be seen in Figure 17a, which shows the spectrum of a film 80 nm in thickness, that as energy increases, the valence band is replaced by an approximately 15 meV indirect band gap in the bulk of the film, inside which there are surface bands of delocalized electron states. Because the thickness of the film is finite, its spectrum in the allowed bulk bands is a collection of dimensional quantization subbands with a small ( 1 meV) distance between them in the valence band and an order of magnitude larger distance in the conduction band.

\begin{figure}[ht]
\includegraphics[width=8cm]{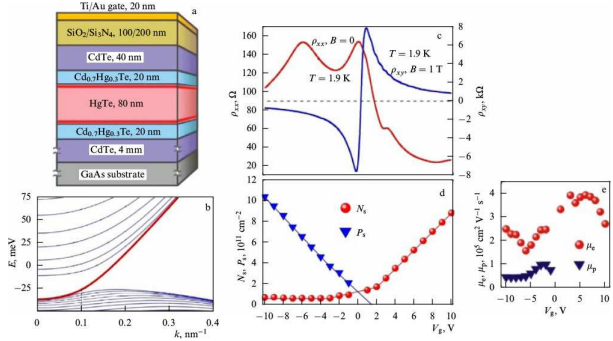}
\caption{(Color online)(a) Cross section of the studied structures. (b) Energy spectrum of a strained 80 nm mercury telluride film. (c) Dependences $\rho_{xx}$ and $\rho_{xy}$. (d) Electron and hole densities as functions of the gate voltage. (e) Average electron and hole mobility.}
\end{figure}

A field-effect transistor structure has been produced based on such a film and is considered in this section. Two types of structures have been studied, whose schematic cross section is shown in Fig. 17a. The structures were grown using molecular beam epitaxy on a (013) oriented CdTe substrate. The bulk of both structures is an 80 nm HgTe layer. The structures differ in the order of the upper layers: the first structure (open) ends with an HgTe layer, while in the second structure ('closed') the main layer is covered with a CdHgTe layer 20 nm in thickness. One of the main achievements of the developed technology, first described in our study [20], is high mobility (up to $5\times10^5 cm^2/Vs$) and low concentration of uncontrolled bulk impurities, which is reduced to $10^{16} cm^{-3}$. This result was achieved due to the use of a 20 nm buffer CdHgTe layer between the HgTe film and the CdTe substrate, which resulted in a sharp decrease in the number of dislocations and defects. This achievement made it possible to obtain not only unambiguous experimental confirmation of the existence of a 3D TI based on a strained HgTe film but also quantitative information on its energy spectrum and on the relative contribution of bulk holes and bulk and surface electrons to the transport response.

\subsection{ Semiclassical transport}

To perform magnetotransport measurements, we made Hall bridges $50 \times 450 \mu m$ in size with a distance of 100 and 250 $\mu m$ between contacts (see the inset in Fig. 3). The central part of the bridges was equipped with a metallic Ti/Au gate. The bridges were produced from both types of structures described in Section 4.1 using standard photolithography and chemical etching. As a gate dielectric, we used either a two-layer film consisting of a 100 nm SiO2 layer and an Si3N4 layer with a thickness of 100–200 nm grown using plasma chemical deposition technology at $T = 100^{\circ} C$, or an 80 nm $Al_2 O_3$ film grown using atomic-layer deposition technology at $T = 80^{\circ}C$. We note that this technology is not much different from that used to produce field-effect transistors based on 2D TIs.

Figure 17c shows a typical dependence of the resistance $\rho_{xx}$  at B = 0 and the Hall resistance $\rho_{xy}$  at B = 1T as a function of the gate voltage $V_g$ at T = 1.9 K. Several maxima are observed on the $\rho_{xx}(V_g)$ curve, the main one being close to $V_g = 1 V$. The curve is asymmetric with respect to the main maximum: the resistance to the left of the maximum is much larger than that to the right, while two side maxima are observed on the curve. The first maximum is located at $V_g = -5.5 V$; its height is the same as that of the main maximum, and the second, much lower, maximum is located at $V_g = 3.5 V$. The $\rho_{xy}$ dependence displayed in the same figure is asymmetric with respect to $V_g = 1 V$, at which it crosses the abscissa axis. A change in the sign of $\rho_{xy}$ suggests that when the gate voltage changes, the Fermi level passes through both the valence and conduction bands. If $V_g > 1 V$, the Fermi level is located in the valence band, where, according to the spectrum shown in Fig. 17b, holes and Dirac surface electrons coexist. This phenomenon is confirmed by the large positive magnetoresistance and the nonlinear Hall effect typical of electron–hole systems.

The dependences of the hole density $P_s$ and mobility $\mu_p$ as well as the total electron density $N_s$ and the average mobility $\mu_e$ on the gate voltage are shown in Fig. 17d, e. These parameters were determined by fitting the calculated dependences $\rho_{xx}(B)$ and $\rho_{xy}(B)$ obtained in the model of classical Drude two-species magnetotransport to the dependences experimentally measured at fixed gate voltages.

We first analyze the behavior of the electron and hole densities. These densities undergo significant variations (by almost an order of magnitude), which indicates a small amount of residual impurities in the film. The CNP is located near the zero gate voltage. The CNP corresponds to the location of the Fermi level near the valence band ceiling, and the densities of bulk holes and surface Dirac fermions in it are the same.

We note once again that the Dirac point does not coincide with the CNP and is experimentally unattainable in our samples, because even at the maximum negative $V_g$ values there is a significant number of electrons in the system, i.e., Dirac electrons contribute to the conductivity at all gate voltages used in the experiment. Fitting based on the Drude model can no longer be considered reliable in the vicinity of the CNP. Therefore, the hole density in the CNP region can only be obtained by extrapolating the dependence $P_{s}(V_{g})$, which crosses the abscissa axis near $V_g = 2 V$. It can be assumed that at this voltage the Fermi level coincides with the valence band ceiling.

Thus, the semimetallic state of the system is realized at $V_g > 2 V$, which emerges as a result of the overlap of the bulk valence band and the surface electron band. For $V_g > 2 V$, there is a small—but the most interesting—voltage region in which transport is determined only by surface electrons (a 3D TI) followed by the start of filling the bulk electron band (the electronic metal state).

We now discuss the behavior of mobility. The hole mobility as a function of the gate voltage is represented by a curve that has a maximum with the mobility value $10^5 cm^2/Vs$, and saturates if $P_s$ increases further. The $\mu_e(V_g)$ dependence is more interesting: there is a wide maximum near $V_g = 5 V$, where $\mu_e(V_g)$ is $4 \times10^5 cm^2/Vs$, which is followed by a minimum at $V_g = -6 V$.
The described behavior of mobility can be associated with both the possible complete exhaustion of one of the surfaces with Dirac electrons (apparently located closer to the gate) and the beginning of the filling of the second hole subband. The valence band ceiling corresponding to the gate voltage $V_g = 2 V$ is confirmed by the temperature dependence$\rho_{xx}(V_g)$  shown in Fig. 18b. It is clearly seen that the point $V_g = 2 V$ is a border-line point: temperature dependence is virtually unobservable to the right of it, while to the left of it resistance significantly increases as the temperature grows. This behavior is associated with the emergence of electron–hole scattering driven by the Landau mechanism [56, 57], similar to that observed in 2D semimetals [58].

\begin{figure}[ht]
\includegraphics[width=8cm]{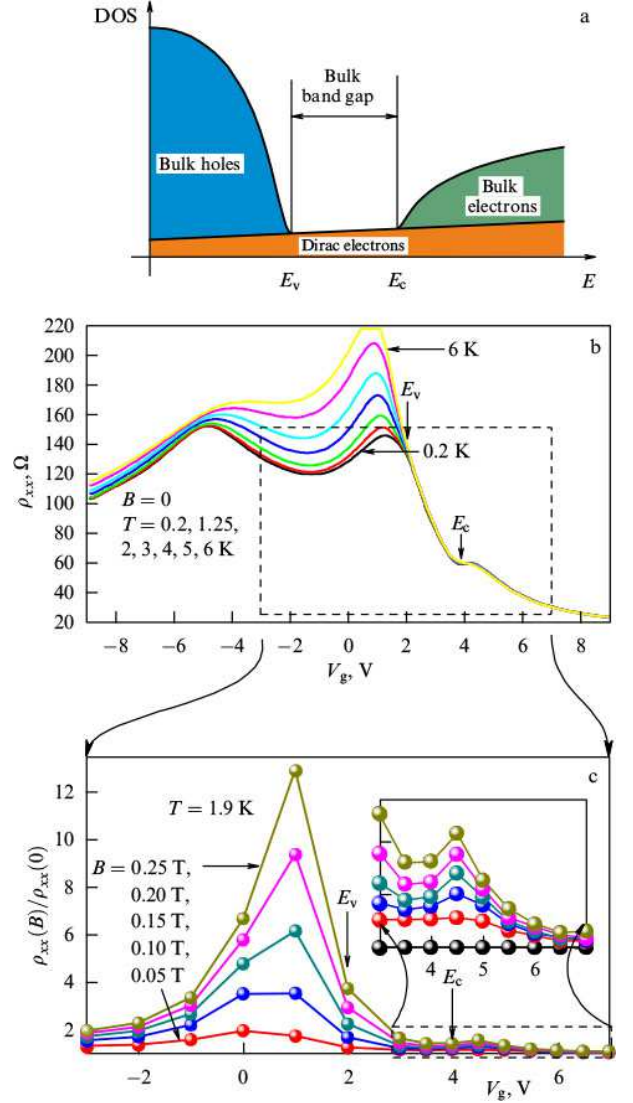}
\caption{(Color online)(a) Representation of the density of states (DOS) as a function of energy. (b) Dependence $\rho_{xx}(V_g)$ at various temperatures in the absence of a magnetic field (vertical arrows indicate the gate voltage values that correspond to the valence band ceiling $E_v$ and the conduction band bottom $E_c$). (c) Dependence of the magnetoresistance amplitude ($\rho_{xx}(B)/\rho_{xx}(B=0)$) in the gate voltage range near the energy band gap indicated by the dashed line (inset shows the region $3 < V_g < 7 V $ on an enlarged scale).}
\end{figure}

This scattering apparently only occurs when the Fermi level crosses the valence band ceiling. Another feature in the $\rho_{xx}(V_g)$ dependence is clearly exhibited at $V_g = 4 V$. Moreover, it blurs at temperatures that exceed 5 K. This feature can be associated with the beginning of the filling of the bulk electron band.

Thus, if the suggested identification of the band boundaries schematically depicted in Fig. 18a is correct, then transport due to surface states is only possible for $2 < V_g < 4 V$. This picture is confirmed by the specific features of classical magnetotransport, more precisely, by the behavior of the dependence of the relative positive magnetoresistance (PMR) ($\rho_{xx}(B)/\rho_{xx}(B=0)$) on the gate voltage.

It should be kept in mind that according to the Drude model, the PMR value in the semimetal state is proportional to the sum of the electron and hole mobilities and, if two groups of electrons coexist, to the difference between the electron mobilities. In accordance with these arguments, the PMR dependence on the gate voltage exhibits a maximum near the CNP, i.e., when the densities of electrons and holes are approximately equal. A rapid decrease (by an order of magnitude) in $\rho_{xx}(B)/\rho_{xx}(B=0)$ is observed to the right of the maximum as the Fermi level moves to the band gap and holes disappear. If $V_g$ increases further, a monotonic decrease occurs that ends near $V_g = 4 V$, and a small maximum, now related to the emergence of bulk electrons, appears of the MR dependence on the gate voltage.

Thus, detailed analysis of the specific feature of classic transport enabled arriving at a self-consistent picture of energy bands. The band gap value found using the difference between electron densities at Vg = 4 V and Vg = 2 V proved to be 15 meV, a value that agrees well with calculations. To conclude, we add that the behavior of the Shubnikov–de Haas (SdH) oscillations presented in Section 4.3 also confirms the described picture.

\subsection{Quantum transport}

In this section, we discuss the specific features of the behavior of SdH oscillations and the quantum Hall effect. Figure 19 displays the dependences $\rho_{xx}(V_g)$ and $\rho_{xy}(V_g)$ measured in magnetic fields up to 4 T. As the magnetic field increases, a sharp increase is observed in the maximum resistance located at the CNP at $V_g = 1 V$, where it reaches $10^7 \Omega/\Box$ at B = 10 T (not shown in the figure). The sign of the dependences $\rho_{xy}(V_g)$ changes at the same gate voltage. A monotonic dependence with minor inflection is only observed with a maximum field of 4 T to the left of the CNP, i.e., in the hole region. In contrast, to the right of the CNP, where the conductivity is determined by highly mobile electrons, well-pronounced QHE plateaus in the $\rho_{xy}(V_g)$ dependence occur already at B = 2T. We note that the QHE also occurs in the $V_g$ regions in which Dirac and bulk electrons coexist. We now analyze this situation in more detail.

\begin{figure}[ht]
\includegraphics[width=8cm]{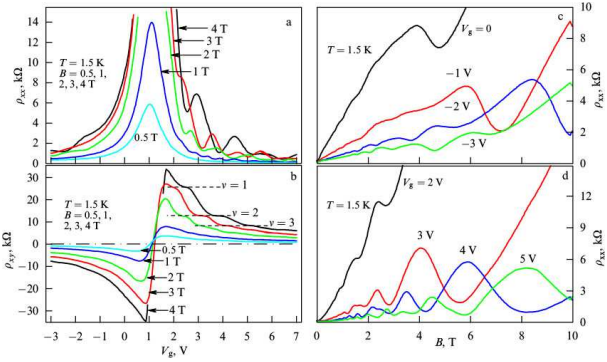}
\caption{(Color online) (a) Dependences $\rho_{xx}(V_g)$ measured in various magnetic fields at T = 1.5 K. (b) Dependences $\rho_{xy}(V_g)$ measured under the same conditions; horizontal dashed lines show the theoretical values $h/(\nu e^2)$ of the corresponding QHE plateaus. Dependences $\rho_{xx}(B)$ for fixed values of $V_g$ for the (c) hole and (d) electron regions.}
\end{figure}

The dependences $\rho_{xx}(B)$ for fixed $V_g$ values are displayed in Fig. 19c, d. The observed picture corresponds as a whole to the dependences on the gate voltage: on the hole side, the SdH oscillations are rather weakly pronounced; on the electron side, on the contrary, deep minima are observed due to high mobility, which indicates that the QHE regime is in effect, and well-pronounced $\rho_{xy}$  plateaus are formed in magnetic fields as low as 2 T.

However, no exponentially small values are observed at the $\rho_{xx}$ minima even in large fields. This observation may be an indication of possible parallel conduction channels, for example, along the lateral surfaces of the film oriented along the applied field. The electron density determined using the position of the SdH oscillation minima in strong magnetic fields ($B > 1-2 T$) turned out to be equal to the density calculated using the Drude model. These values are compared in Fig. 20d. It follows from the coincidence of and that the filling factors $\nu$ are determined by the total density , i.e., the sum of the densities of Dirac and bulk electrons. Similarly, the hole densities obtained for large negative Vg from the analysis of SdH oscillations in strong fields and from fitting in the Drude model turn out to be quite close. However, in approaching the CNP, is systematically smaller than . It can be concluded based on these observations that the behavior of the QHE, when the Fermi level is located in the valence band, is driven by the difference between the densities of holes and electrons. The essentially important conclusion is that in strong magnetic fields, surface charge carriers also participate in the formation of unified Landau levels.

\begin{figure}[ht]
\includegraphics[width=8cm]{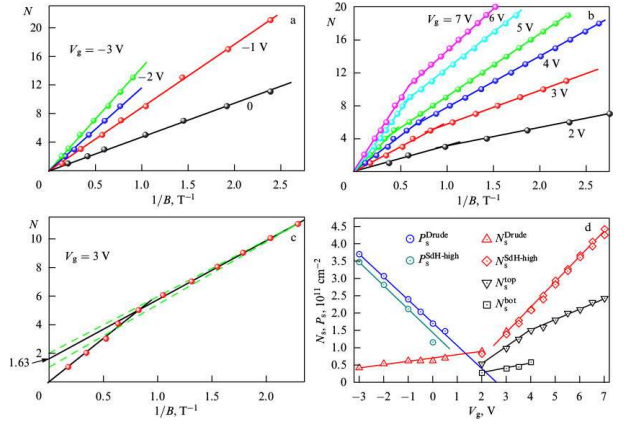}
\caption{(Color online)Number N of the oscillation minima determined from the dependences $\rho_{xx}(B)$  shown in Fig. 19c, d as a function of 1/B for (a) hole and (b) electron regions. (c) Dependences N(1/B) at $V_g = 3 V$; solid black lines correspond to the best fit of the strong- and weak-field parts of the dependence; the arrow at the ordinate axis indicates the point of intersection with the vertical axis; dashed lines show possible fittings subject to the condition that the vertical axis is crossed at a point with an integer value. (d) Comparison of the electron $N_s$ and hole $P_s$ densities determined in various ways: by fitting the dependences
$\rho_{xx}(B)$ and $\rho_{xy}(B)$ in the Drude model and by analyzing the position of the minima of SdH oscillations in strong magnetic fields; the density of Dirac electrons on the upper surface of is determined from an analysis of SdH oscillations in weak fields, and on the lower surface,as the difference between and. }
\end{figure}

To analyze the behavior of the SdH oscillations even more deeply, we determined how the number N of the minima of these oscillations shown in Fig. 19c, d depends on their position on the axis of the inverse magnetic field 1/B. These dependences are plotted in Fig. 20a, b. The oscillations are weakly pronounced on the hole side (Fig. 19a, c). In magnetic fields less than 1–2 T, only oscillations with odd numbers remain discernible with magnetic fields up to B = 0.4 T, with corresponding filling factors exceeding 10. Each of the obtained dependences is well approximated by a straight line passing through the origin. The slope of this line corresponds to the differential hole–electron concentration mentioned above.

The SdH oscillations on the electron side are much more pronounced (Fig. 19d), regardless of whether the Fermi level is located in the band gap or in the conduction band. The oscillations remain discernible in magnetic fields up to 0.25 T with corresponding filling factors greater than 20. A more careful analysis shows that regions of weak and strong magnetic fields with a sharp transition between them can be found in any of the dependences displayed in Fig. 20b. The periodicity of oscillations as a function of the inverse magnetic field persists in each of the regions; however, the slopes of the lines that correspond to these regions of the N(1/B) dependence not the same. The difference between the densities found using the slope of these lines is $20–45\%$.

Thus, the presence of two regions with different slopes cannot be explained by any degeneracy lifted by the magnetic field.

On the other hand, the presence of two densities, determined by the periodicity of SdH oscillations in weak and strong magnetic fields, can be explained by the existence of two (or more) groups of carriers, each of which has its own set of Landau levels. Such a situation is quite possible if the effects of gate screening by the upper surface are taken into account. This would result in different densities of Dirac electrons on the upper and lower surfaces.

We now assume that flat bands in the system under study are formed near the zero gate voltage, and the concentrations and are equal. The band diagram of the structure under study at this gate voltage and at other $V_g$ is schematically depicted in Fig. 21. The flat-band situation is actually realized if $N_{s}^{top}=N_{s}^{bot}$. However, as the gate voltage increases, the density $N_{s}^{top}$ increases much faster than $N_{s}^{bot}$. The ratio of the filling rates $\alpha=(dN_{s}^{top}/dV_g)/(dN_{s}^{bot}/dV_g)$ of the surfaces can be estimated in the simplest way in the absence of bulk carriers, i.e., at $2 \leq V_g \leq 4 V$. The ratio is given under these conditions by the formula $\alpha = 1 + (e^2Dd_{HgTe}/\epsilon_{HgTe}\epsilon_{0})$, where D is the density of states of Dirac electrons on the upper surface and $d_{HgTe}$ and $\epsilon_{HgTe}$ are the thickness and dielectric constant of the mercury telluride film. Substituting typical values, we obtain $\alpha = 3-5$.

\begin{figure}[ht]
\includegraphics[width=8cm]{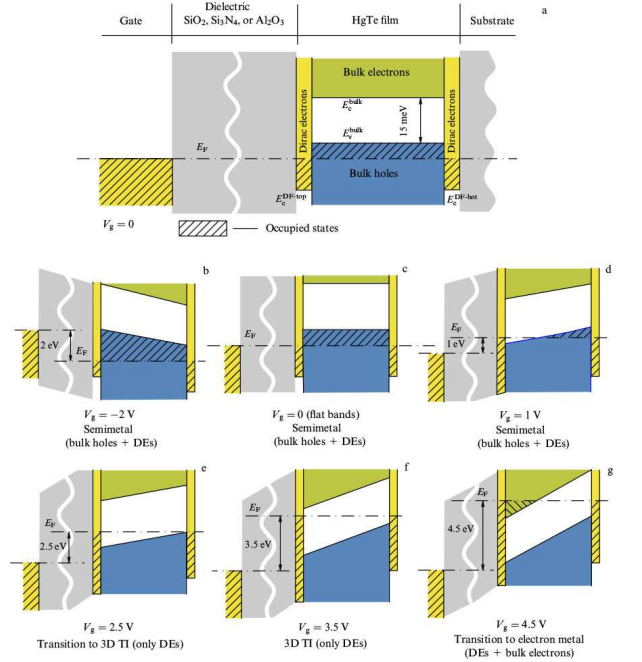}
\caption{(Color online)Schematic representation of the band diagram of the system under study at various gate voltages (the voltage drop in the dielectric is shown not to scale). Dirac points for both surfaces denoted as and are located in the valence band. Flat bands correspond to a zero voltage at the gate. Application of a gate voltage leads to the bending of the bands; the distant surface (bottom) is partially screened by the nearer one (top) and bulk carriers. In the range $2 < V_g < 4V$, the Fermi level is located in the band gap (e, f), while for $V_g < 2 V$, the system contains bulk holes (a–d), and for $V_g > 4 V$, bulk electrons (g)}
\end{figure}

Thus, the emergence of an electric field in the HgTe film should lead to electron densities that are different on the top and bottom surfaces. The main mechanism of electron scattering at T = 4.2 K is their scattering by residual impurities. It is then natural to assume that, of two identical groups of carriers, the higher-density group has greater mobility and less broadening of the Landau levels. The Landau levels on the top surface are less broadened in this case, and SdH oscillations start forming for this surface in weaker fields. As a result, the SdH oscillation period in weak fields only yields the electron density on the top surface, while both groups of carriers (to which bulk electrons are added at $V_g > 4 V$) are quantized in strong fields, and the oscillation period yields their total density.

The dependence determined in this way is shown in Fig. 20d. Because only surface electrons are present in the system in the range $2 \leq V_g \leq 4 V$, the relation is valid in this range, and can be determined. As expected, the experimentally found filling rate is three times lower than $dN_{s}^{top}/dV_g$. Another specific feature should be emphasized: a sharp bending of the dependence at $V_g = 4 V$. A decrease in the slope $dN_{s}^{top}/dV_g$ at $V_g > 4 V$ apparently can only be associated with a decrease in the contribution of the density of states of surface electrons to the total density of states. This conclusion is in line with the previously proposed picture of the energy spectrum in which the Fermi level enters the conduction band at the indicated gate voltage.

Finally, the assumption that the weak-field part of the SdH oscillations emerges due to Dirac fermions is confirmed by the phase of these oscillations, the analysis of which for a system in the TI state ($V_g = 3 V$) is presented in Fig. 20. The formula $(1/B_{min})/\Delta_{1/B} = \nu_{tot}$ turns out to be correct for the strong-field part of the dependence. Here, $1/B_{min}$ is the location of the minima in the inverse magnetic field, $\Delta_{1/B} $ is the period of oscillations in the inverse field determined by the total density, and $\nu_{tot}$ is an integer that corresponds to the total filling factor for all types of electrons. However, if the weak-field part of the $N(1/B_{min})$ dependence is approximated by a linear function and continued until it crosses the vertical axis, then the intersection occurs at 1.63. The obtained linear dependence is described by the formula
$(1/B_{min})/\Delta_{1/B} = \nu_{top}+0.63$, where $\nu_{top}$ is the electron filling factor on the top surface (determined up to an integer), and 0.63 is the phase oscillation shift. An approximation of the weak-field part by a linear dependence with the phase shift ignored (dashed lines in Fig. 20c) yields an inferior result. Thus, a phase shift of $0.63 \pm 0.023$ is observed in weak-field oscillations, which is close to the predicted value of 0.5 for spin-polarized Dirac fermions [59].
\subsection{ Capacitive spectroscopy of Shubnikov–de Haas oscillations}

As can be seen from Section 4.3, an analysis of SdH oscillations indicates the presence of their anomalous phase. However, because the transport response contains competing contributions from both surfaces of the TI, this does not allow drawing an unambiguous conclusion that the transport oscillations of the top surface are undisturbed by the contribution of the bottom one. Therefore, a conclusion on the anomalous phase being observed can only be made with some caution.

We show in this section that the capacitive spectroscopy of a 3D TI makes it possible to circumvent this difficulty and obtain more accurate and detailed information on the behavior of the SdH oscillations of 2D DFs on one (upper) surface and thus demonstrate the anomalous behavior of their phase due to rigid topological coupling of spin and momentum.

Figure 22 shows the structure of the sample under study and the equivalent circuit in which capacitive response is measured. The circuit diagram clearly shows that the capacitance of the upper DFs is separated from that of the lower DF by the capacitance of the film and therefore the capacitive response of the upper surface of the investigated TI should 'feel' the effect of the lower one to a much lesser extent than when the transport response is measured. Detailed measurements of the capacitive SdH oscillations have confirmed this assumption. The measurements showed that when the Fermi level is located in the TI band gap, capacitive oscillations of only the upper DFs are observed, which exhibit an anomalous phase in a wide range of magnetic fields. Below, we discuss these observations in detail.

\begin{figure}[ht]
\includegraphics[width=8cm]{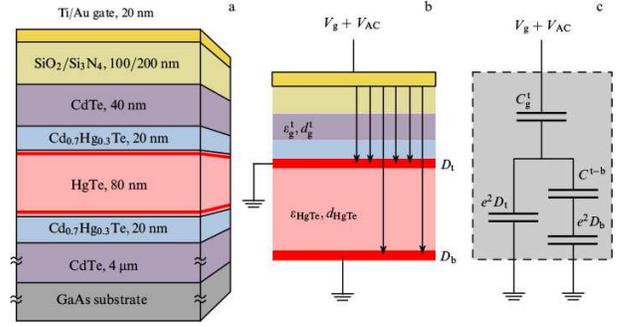}
\caption{(Color online) (a) Section of the structure under study. (b) Electric circuit for connecting layers to a gate voltage source and the pattern of electric field lines. (c) Equivalent circuit of structure capacitance. VAC is the measuring variable signal, is the dielectric constant of the insulator layer between the gate and the upper surface of the HgTe film, is the thickness of this layer, $\epsilon_{HgTe}$ is the dielectric constant of HgTe, dHgTe is the thickness of the HgTe film, is the capacitance between the gate and the upper surface of the HgTe film, $C_{tb}$ is the capacitance between the top and bottom surfaces of the HgTe film, $D_t$ is the density of DF states on the upper surface, $D_b$ is the density of DF states on the bottom surface.}
\end{figure}

Figure 23b displays the capacitance $C(V_g)$ as a function of the gate voltage in the absence of a magnetic field and at B = 2T. It can be seen that in the absence of the magnetic field, $C(V_g)$ has a minimum, which, as a comparison with transport data shows (Fig. 23a) corresponds to the Fermi level passing through the bulk gap, while the application of a magnetic field results in the emergence of SdH oscillations. It is clearly seen that the oscillations are virtually absent when the Fermi level is located in the valence band and have a significant amplitude when it passes through the bulk gap and the conduction band. Such behavior shows that the DF mobility significantly increases when the Fermi level exits the valence band.

\begin{figure}[ht]
\includegraphics[width=8cm]{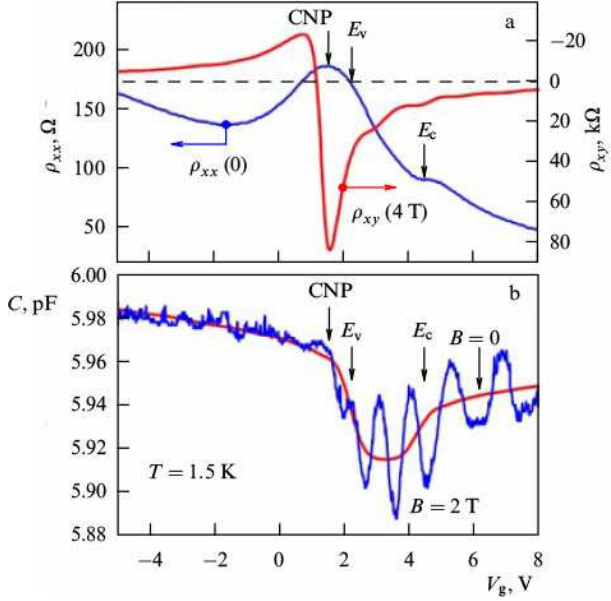}
\caption{(Color online)a) Dependences $\rho_{xx}(V_g)$ and $\rho_{xy}(V_g)$. (b) Capacitance as a function of the gate voltage in a zero magnetic field and in a 2 T field. }
\end{figure}

An analysis of the period of oscillations inside the band gap shows that they are determined for all magnetic fields by nondegenerate Landau levels of the DFs on the upper surface. This is evidence that, as one would expect, 2D DFs of the upper surface form a spin-polarized system, thus indicating its topological nature. Vivid evidence of this nature is the behavior of the SdH oscillation phase [59] described by the formula $\frac{1/B_{min,n}}{\Delta_{1/B}}=n+\delta$

where $B_{min}$, n is the location of the n-th minimum of oscillations and $\Delta 1/B$ is the period of oscillations in the inverse magnetic field. We can use this formula to determine the phase by linearly extrapolating the oscillation period to zero on the scale of the inverse magnetic field 1/B as a function of the oscillation number n. Due to the selectivity of capacitive spectroscopy noted above, which allows studying the DF oscillations of only the upper surface of the TI, the required set of oscillations can be obtained with high accuracy in a wide range of magnetic fields by measuring the dependences C(B) for given $V_g$.

Figure 24a shows the dependences of 1/B on n determined in this way. Shown for comparison in the same figure is a similar dependence but plotted based on the analysis of magnetotransport data at $V_g$ = 6 V, when the Fermi level is located deep in the conduction band. It is clearly seen that while the minima obtained from the analysis of the capacitance are well described by straight lines crossing the horizontal axis with the expected offset from zero, the corresponding dependence of the transport oscillation minima can only be described using two straight lines that separate the obtained dependence into weak-field and strong-field regions. In weak fields, where the magnetotransport oscillations only emerge due to the upper surface and thereby reproduce the behavior of the magnetic capacitance, the phase shift obtained from the transport data $\Delta_{transport} = 0.72 \pm 0.04$ yields the same value that was found from the capacitance data, $\Delta_{capacitance} = 0.7 \pm 0.04$. More than one carrier group is involved in strong fields in the formation of transport oscillations, and the shift extracted from data fitting in this region is already close to zero, as is expected for a conventional 2D electron system. Figure 24b shows the evolution of the phase $\delta$, extracted from the capacitance data, for all positions of the Fermi level. It is clearly seen that $\delta$ is close to 0.5 just when the Fermi level is located in the bulk gap, i.e., when capacitance oscillations are only formed by topologically stable 2D DFs of the upper surface. As soon as $E_F$ enters the valence band, the phase rapidly vanishes and, in contrast, gradually increases as $V_g$ increases, approaching unity at the maximum positive $V_g$. This observation can be explained by the hybridization of surface and bulk carriers deep in the conduction band.

\begin{figure}[ht]
\includegraphics[width=8cm]{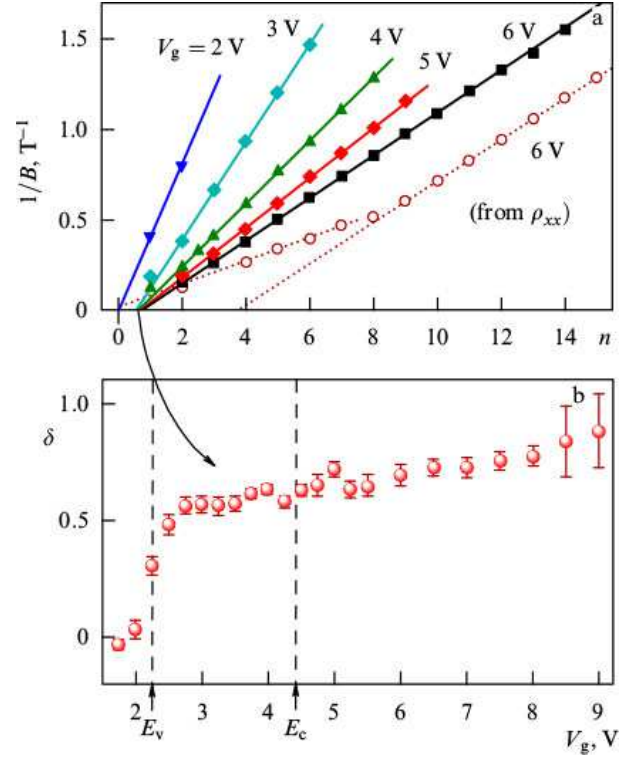}
\caption{(Color online)(a) Positions of the minima of magnetocapacitance and magnetoconductivity oscillations (dependence with a kink) in the inverse magnetic field, measured at fixed gate voltages $V_g$ = 2, 3, ..., 6 V as a function of oscillation number n. Drawn through the experimental dependences are fitting straight lines extrapolated to the intersection with the horizontal axis. The point where the lines cross the axis corresponds to the oscillation phase $\delta$ associated with the Berry phase of Dirac electrons. (b) Phase $\delta$ obtained from the data displayed in (a) as a function of the gate voltage.}
\end{figure}

\section{Conclusion}

We did not aim in this review to provide the widest possible presentation of studies related to TIs based on HgTe. On the contrary, we rather tried to formulate the most important and fundamental facts that any specialist that begins studying topological HgTe insulators needs to know. Nevertheless, we should primarily mention studies of photovoltaic effects in these insulators [60–62], in particular, the observation of the generation of chiral-spin photocurrents in 2D TIs [61]. We also mention studies on terahertz magnetic spectroscopy of 3D TIs in which the effective masses of surface DFs were measured for the first time [63, 64]. Finally, we note the studies of magnetooptical [65, 66] and magnetotransport [67–69] properties of double CdHgTe/HgTe/CdHgTe heterostructures, which are of importance for understanding the behavior of TIs.

In conclusion, we emphasize once again that HgTe is the only material that allows realizing both 2D and 3D TIs. The results presented show that transport and photoelectric responses reflect all the features associated with the main properties of the TI: the presence of surface states (edge one-dimensional in the case of a 2D TI and surface two-dimensional in the case of a 3D TI) and a rigid topological coupling of the electron spin and momentum. In the case of a 2D TI, this is primarily the existence of nonlocal transport in both ballistic and diffusion regimes.

The most important theoretical prediction regarding topological protection against backscattering in 2D TIs in experimental samples, strictly speaking, has not been observed: the accuracy of quantization of ballistic resistance does not exceed $10\%$, and the mean free path is several micrometers.

A 3D TI based on a strained HgTe film is the cleanest among all 3D TIs known so far. The mobility of Dirac electrons in it is $5 \times10^5 cm^2/Vs$. This feature enabled the determination of all its main parameters: the bulk gap, the DF density on the upper and lower surfaces of the TI, and their effective mass. This also allowed reliably establishing the presence of the Berry phase in SdH oscillations, whose existence reflects the most important property of the TI : the rigid coupling of the electron spin and momentum.

There are a number of unresolved issues associated with the physics of TIs and newly emerged problems that can be solved by studying TIs based on HgTe. We list some of these problems: (1) determining the role of topological protection in kinetic processes (nonequilibrium phenomena, noise, localization) [70, 71]; (2) determining the optical properties of TIs [55]; (3) exploring the properties of hybrid TI-based systems [72, 73]; (4) establishing the properties of nanostructured TI-based systems [74, 75].

Thus, further exploration of topological insulators based on HgTe is of unquestionable interest.

This study was supported by the Russian Science Foundation (grant no. 16-12-10041-P).

\end{document}